\documentclass{camilo_polarimetry_emulateapj}
\usepackage{apjfonts}
\usepackage{psfig}

\journalinfo{The Astrophysical Journal, Letters (in press)}
\slugcomment{Received 2007 January 22; accepted 2007 February 22}

\shortauthors{CAMILO ET AL.}
\shorttitle{POLARIMETRY OF THE MAGNETAR XTE~J1810--197}

\begin{document}


\def\xte{XTE~J1810--197}

\title{Polarized radio emission from the magnetar XTE~J1810--197}

\author{F.~Camilo,\altaffilmark{1}
  J.~Reynolds,\altaffilmark{2}
  S.~Johnston,\altaffilmark{3}
  J.~P.~Halpern,\altaffilmark{1}
  S.~M.~Ransom,\altaffilmark{4}
  and W.~van~Straten\altaffilmark{5}
}

\altaffiltext{1}{Columbia Astrophysics Laboratory, Columbia University,
  550 West 120th Street, New York, NY 10027.}
\altaffiltext{2}{Australia Telescope National Facility, CSIRO, Parkes
  Observatory, PO Box 276, Parkes, NSW 2870, Australia.}
\altaffiltext{3}{Australia Telescope National Facility, CSIRO, PO Box 76,
  Epping, NSW 1710, Australia.}
\altaffiltext{4}{National Radio Astronomy Observatory, 520 Edgemont Road,
  Charlottesville, VA 22903.}
\altaffiltext{5}{Center for Gravitational Wave Astronomy, The University
  of Texas at Brownsville, TX 78520.}

\begin{abstract}
We have used the Parkes radio telescope to study the polarized emission
from the anomalous X-ray pulsar \xte\ at frequencies of 1.4, 3.2,
and 8.4\,GHz.  We find that the pulsed emission is nearly 100\% linearly
polarized.  The position angle of linear polarization varies gently across
the observed pulse profiles, varying little with observing frequency or
time, even as the pulse profiles have changed dramatically over a period
of 7 months.  In the context of the standard pulsar ``rotating vector
model,'' there are two possible interpretations of the observed position
angle swing coupled with the wide profile.  In the first, the magnetic
and rotation axes are substantially misaligned and the emission originates
high in the magnetosphere, as seen for other young radio pulsars, and the
beaming fraction is large.  In the second interpretation, the magnetic
and rotation axes are nearly aligned and the line of sight remains in
the emission zone over almost the entire pulse phase.  We deprecate
this possibility because of the observed large modulation of thermal
X-ray flux.  We have also measured the Faraday rotation caused by the
Galactic magnetic field, $\mbox{RM} = +77$\,rad\,m$^{-2}$, implying an
average magnetic field component along the line of sight of $0.5\,\mu$G.
\end{abstract}


\section{Introduction}\label{sec:intro} 

Anomalous X-ray pulsars (AXPs) and soft-gamma repeaters are neutron stars
many of whose attributes at X-ray, gamma-ray, and infrared wavelengths
\citep[see][for a review]{wt06} are best understood in the context of
the magnetar model \citep{dt92a}, according to which their high-energy
emission results from the rearrangement and decay of ultra-strong
magnetic fields.  Much remains to be learned about magnetars, of which
only a dozen are known.

\xte\ is an AXP with spin period $P=5.54$\,s, unusual in being transient.
Identified in early 2003 when its X-ray luminosity increased 100-fold
\citep{ims+04}, by 2007 it has returned to the quiescent state it
maintained for at least 24 years \citep{gh05}.  Uniquely for a magnetar it
emits radio waves \citep{hgb+05}, that turned on by early 2004.  Unlike in
ordinary rotation-powered pulsars, the radio pulses have a flat spectrum
and vary in luminosity and shape on daily timescales \citep{crh+06}.

Radio emission from \xte\ links magnetars and ordinary pulsars, and
provides a new window for learning about the physical characteristics
of a magnetar.  For instance, while in principle radio emission could
be generated in the corona from closed or open magnetic field lines, the
large pulse profile and flux density changes observed on short timescales
\citep{ccr+07} appear to point to the latter \citep[cf.][]{bt07}.

Here we report on observations of the polarized emission from \xte\
in an attempt to shed some light on the geometry of the radio-emitting
regions of this magnetar.

\section{Observations and Analysis}\label{sec:obs} 

We have observed \xte\ with the Parkes 64-m telescope in New South Wales,
Australia, in full-Stokes polarimetry mode for a total of 20\,hr on-source
between 2006 April and November.  Table~\ref{tab:obs} summarizes the
relevant observations.

\begin{deluxetable}{llclr}
\tablewidth{0.86\linewidth}
\tablecaption{\label{tab:obs} Parkes polarimetric observations of \xte\  }
\tablecolumns{5}
\tablehead{
\colhead{Date}           &
\colhead{Frequency}      &
\colhead{Integration}    &
\colhead{Backend}        &
\colhead{$S_{\rm peak}$} \\
\colhead{(MJD/mmdd)}     &
\colhead{(GHz)}          &
\colhead{(hr)}           &
\colhead{}               &
\colhead{(mJy)}                 
}
\startdata
53852/0427 & 1.369                  & 0.9 & WBC\tablenotemark{a} &  650 \\
53862/0507 & 1.369                  & 0.2 & WBC\tablenotemark{a} &  350 \\
53879/0524 & 1.369                  & 0.2 & WBC\tablenotemark{a} &  320 \\
53913/0627 & 1.369\tablenotemark{b} & 2.0 & DFB                  & 1500 \\
53986/0908 & 1.369                  & 1.2 & DFB                  &   70 \\
53989/0911 & 8.356\tablenotemark{c} & 5.0 & WBC\tablenotemark{d} &   45 \\
53993/0915 & 3.222                  & 0.3 & DFB\tablenotemark{d} &   20 \\
54002/0924 & 1.369                  & 3.2 & DFB                  &   50 \\
54021/1013 & 1.369                  & 0.4 & DFB\tablenotemark{d} &   20 \\
54021/1013 & 3.222                  & 1.8 & DFB\tablenotemark{d} &   10 \\
54022/1014 & 3.222                  & 3.3 & DFB                  &   20 \\
54060/1121 & 1.369\tablenotemark{b} & 1.6 & DFB                  &   20 \\
\enddata
\tablecomments{Observations at 1.4\,GHz used the central beam of the
multibeam receiver at a variety of feed angles, unless otherwise noted.
Observations at 3.2\,GHz were performed with the 10/50\,cm receiver,
and 8.4\,GHz observations used the Mars receiver.  We used the wide-band
correlator (WBC) or the digital filterbank (DFB) spectrometer.  The
integration times were divided into scans interspersed with calibration
observations.  Scans were divided into subscans where an integer number
of pulsar periods (minimum of two) were folded modulo the pulsar period.
Unless noted otherwise, the total bandwidth recorded was 256\,MHz (with
128, 512, or 1024 channels across it), and there were 2048 phase bins
across each folded pulse profile (2.7\,ms resolution).
}
\tablenotetext{a}{1024 phase bins.}
\tablenotetext{b}{H-OH receiver.}
\tablenotetext{c}{512\,MHz bandwidth.}
\tablenotetext{d}{Data folded at half the pulse period.}
\end{deluxetable}

We have used the three available receiver/feed combinations that have
well-characterized polarization properties: H-OH (1.4\,GHz), 10/50\,cm
(3.2\,GHz), and Mars (8.4\,GHz).  Due to common availability, at 1.4\,GHz
we have also used the central beam of the multibeam receiver, although
its polarimetric characteristics are less ideal compared to those of
the H-OH receiver \citep{joh02}.  The H-OH and 10/50\,cm systems have
orthogonal linear feeds, while the Mars package receives dual circular
polarizations.  In all cases a pulsed calibrating signal can be injected
at an angle of 45\arcdeg\ to the feed probes.

To record data we used either the digital filterbank (DFB) or the
wide-band correlator (WBC).  The bandwidth, frequency- and time-resolution
varied depending on receiver and spectrometer, but typical values were,
respectively, 256\,MHz, 128 channels, and 2048 bins across the pulse
profile (see Table~\ref{tab:obs} for details).  An integer number of
pulse periods were folded and recorded to disk in PSRFITS format for
off-line analysis.  Because the dump time of the spectrometers is $\ge
10$\,s, a minimum of two pulse periods were folded in each subscan,
with $\sim 10$ more common.  Typically, scans lasting up to $\sim
1$\,hr were interspersed with $\sim 1$\,min observations of the pulsed
calibrator in order to determine the relative gain and phase between
the two feed probes.  For our purposes, the main difference between
the 3-level sampling/correlation WBC and the 8-bit precision DFB was
the latter's much greater sensitivity to radio frequency interference,
which we excised in the frequency- and time-domain during analysis.

We used existing observations of the flux calibrator Hydra~A, whose flux
density is 43.1, 20.3 and 8.4\,Jy at 1.4, 3.2 and 8.4\,GHz respectively,
to determine the system equivalent flux density for the receivers and
to flux-calibrate the pulse profiles.

All data were analyzed with the {\sc psrchive} software package
\citep{hvm04}.  As part of the analysis we corrected the Stokes parameters
($Q$, $U$, total intensity $I$, and circular polarization $V$) for
the position of the feed probes relative to the telescope meridian and
for the parallactic angle of the observation.  We also observed strong
pulsars with known polarization characteristics (such as the Vela pulsar)
to provide a check on our polarimetric calibration.  Analysis of these
pulsars yielded linear polarization $L = \sqrt{Q^2 + U^2}$, position
angle of linear polarization $\mbox{PA} = \frac{1}{2} \arctan\ (U/Q)$,
and $V$ matching those in the literature \citep[e.g.,][]{jhv+05,jw06}.

\section{Results and Discussion}\label{sec:res} 

\begin{figure}
\begin{center}
\includegraphics[angle=0,scale=1.22]{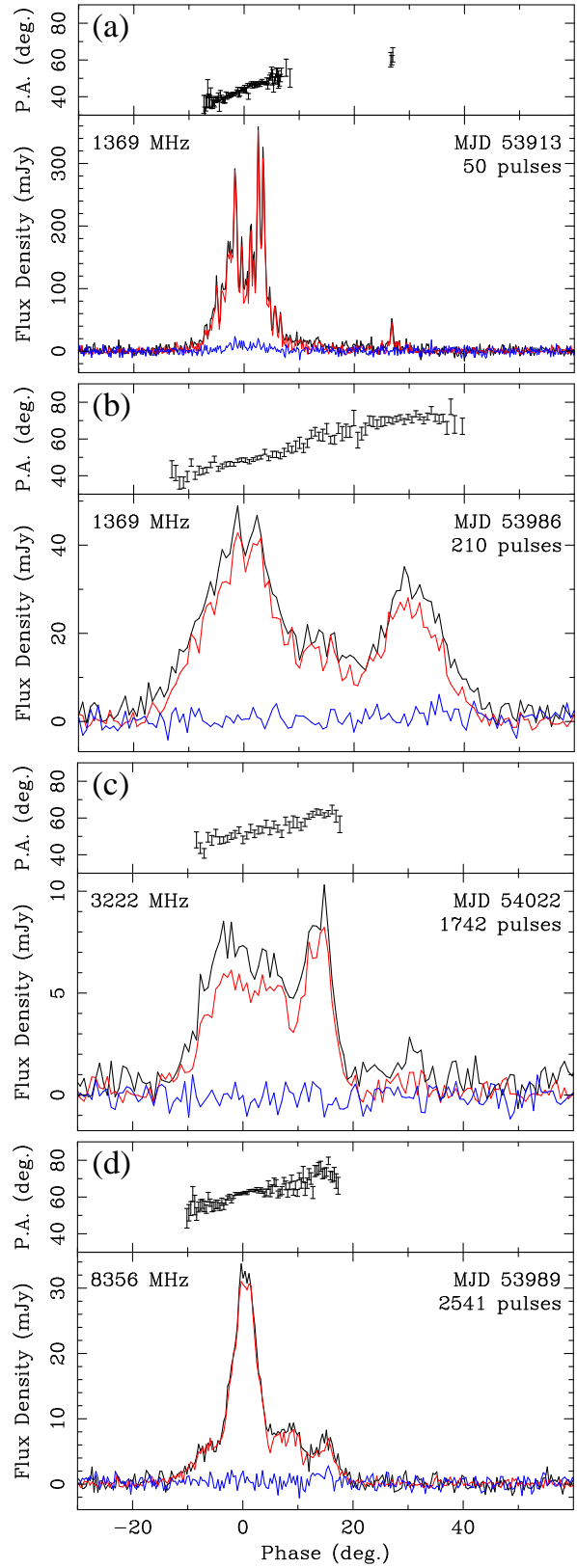}
\caption{\label{fig:pol}
Representative polarimetric profiles of \xte\ over time and frequency.
The \textit{red} and \textit{blue} lines represent the amount of
linear and circular polarization, respectively, and the \textit{black}
traces give total intensity.  The position angles (PAs) are displayed
for bins where the linear polarization signal-to-noise ratio $> 4$
and have been corrected for rotation measure.  In order to highlight
the emission region, we show 90\arcdeg\ of pulse phase and a PA range
of 30\arcdeg--90\arcdeg.  Each data set was folded using a different
ephemeris, and no attempt was made to align the panels precisely.
Panel (a) shows a 5\,min integration with 2048-bin resolution when the
pulsar was strong (emission on this day was also detected at other phases;
see Fig.~\ref{fig:sp}).  Panel (b) displays a 19\,min integration when
the pulsar had weakened and its profile had broadened considerably.
Panel (c) shows a 2.7\,hr integration, and (d) a 3.9\,hr integration.
Panels (b)--(d) are displayed with 512 bins across the full pulse period.
}
\end{center}
\end{figure}

To complete polarization calibration for \xte\ we had to compute the
amount of Faraday rotation suffered by the radiation in its passage
through the Galactic magnetic field.  We determined the rotation measure
by measuring PA as a function of frequency within the 256-MHz band at
1.4\,GHz when the pulsar was strong.  The resulting value, $\mbox{RM} =
+76 \pm 4$\,rad\,m$^{-2}$, did not vary within the quoted uncertainty
either as a function of pulse phase or time. The RM was then used to
correct the measured PAs and frequency-integrated $L$ at all frequencies
to their values at infinite frequency so that a comparison could be made
between frequencies \citep[e.g.,][]{kj06}.  The PAs and $L$ shown in
Figure~\ref{fig:pol} therefore represent those emitted at the pulsar.
We also display in the Figure the Stokes $I$ and $V$.

Together with the integrated column density of free electrons to \xte,
$\mbox{DM}=178$\,cm$^{-3}$\,pc \citep{crh+06}, the RM can be used to
determine the average magnetic field strength parallel to the line of
sight weighted by electron density, $1.2\,\mbox{RM/DM} = 0.5\,\mu$G.  This
fairly small value appears reasonable given the location of the pulsar,
$(l,b) = (10\fdg73, -0\fdg16)$ and $d=3.5$\,kpc, for which the large-scale
Galactic field is mostly in the perpendicular direction, and with at
least one reversal along the line of sight \citep[e.g.,][]{hml+06}.

\begin{figure}
\begin{center}
\includegraphics[angle=0,scale=1.70]{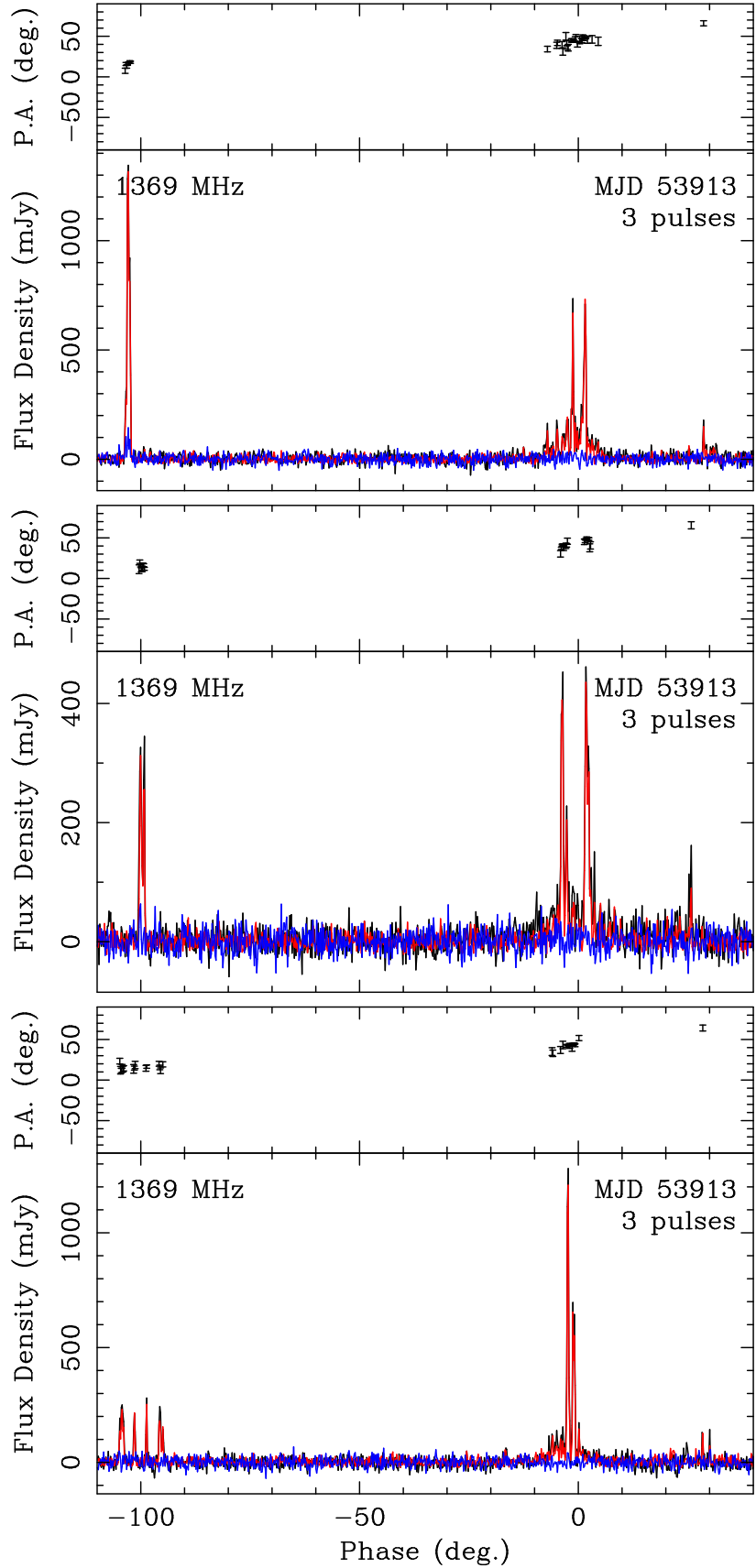}
\caption{\label{fig:sp}
Selected polarimetric profiles of \xte\ at 1.4\,GHz on MJD~53913
(see also Fig.~\ref{fig:pol}~[a] for data collected 6\,hr earlier).
Each panel contains 16.6\,s of data (three pulses) folded modulo the
pulsar period.  The top two panels are contiguous in time, with the one
at bottom recorded 6\,min later.  In order to highlight the emission
region, we show in each panel 150\arcdeg\ of pulse phase.  Traces and
symbols are as in Figure~\ref{fig:pol}.
}
\end{center}
\end{figure}

In spite of the variability of the profiles shown in Figure~\ref{fig:pol}
there are three striking and constant aspects to the polarization
profiles.  First, the fractional linear polarization is extremely
high, close to 100\%, and remains high at all frequencies measured
here\footnote{\citet{crh+06} reported that the pulsar was 65\% linearly
polarized at 1.4\,GHz.  The discrepancy arises from then-uncorrected
Faraday rotation within the observing band.}.  Secondly, there is a
shallow increase in the PA as a function of rotational phase, which
remains essentially unchanged regardless of time or frequency of the
observations.  The rate of change is reasonably constant over the ``main''
pulse profile components and is around $0\fdg5$\,deg$^{-1}$.  Finally,
there is little or no circular polarization ($\la 5\%$) in the integrated
profiles at any frequency (Fig.~\ref{fig:pol}), or in individual pulses
at 1.4\,GHz except for occasional levels up to $\sim 10\%$ of total
intensity in the ``precursor'' pulse components (cf. Fig.~\ref{fig:sp}).

The emission from \xte\ changed in character in late 2006 July
\citep{ccr+07}.  While daily variations continue unabated, generally
the pulse profiles are broader (compare Figs.~\ref{fig:pol}~[a]
and [b]) and the fluxes are lower (the peak flux densities listed
in Table~\ref{tab:obs} attest to this).  In contrast, the general
polarization characteristics do not seem to vary.  This suggests that the
gross observed changes in profile morphology are not due to detectable
changes in the underlying magnetic field geometry of the emission regions.

\begin{figure}
\begin{center}
\includegraphics[angle=270,scale=0.54]{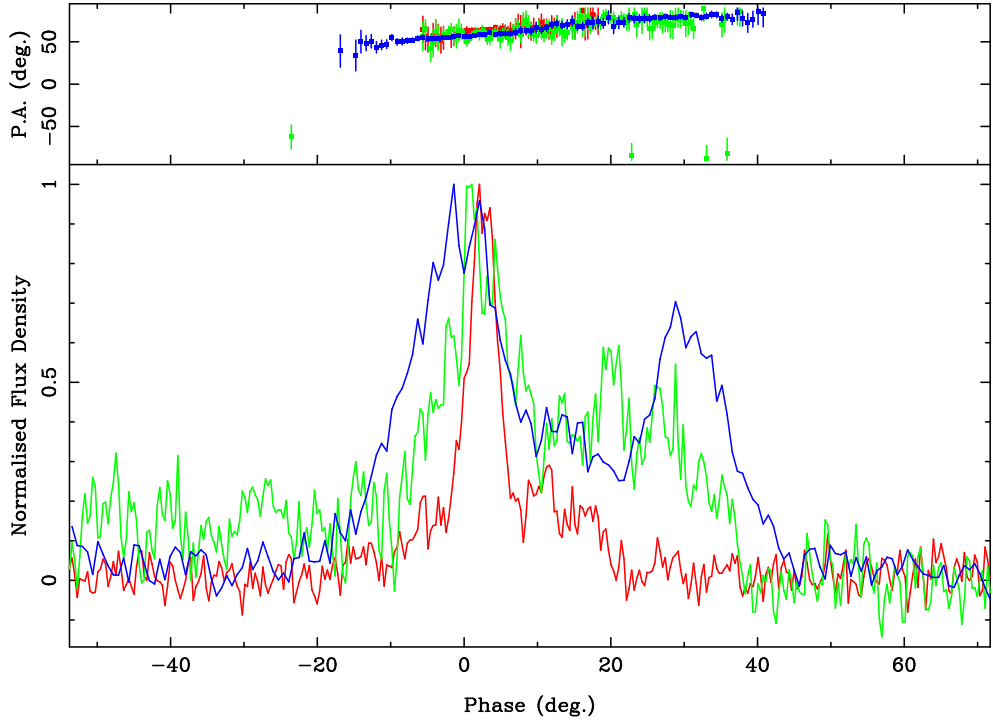}
\caption{\label{fig:pa}
Position angles (\textit{top}) and total intensity profiles
(\textit{bottom}) for \xte\ at three frequencies (\textit{blue}
corresponds to 1.4\,GHz data, \textit{green} to 3.2\,GHz, and \textit{red}
to 8.4\,GHz).  All data were obtained within a one-week period, and
125\arcdeg\ of phase are displayed with 512 bins across the full pulse
period.  The 1.4\,GHz and 8.4\,GHz data were presented in a somewhat
different form in Figure~\ref{fig:pol}, while those at 3.2\,GHz (195
pulses, or 18\,min) are from MJD~53993.  Starting with the previous
best determination of RM ($+76 \pm 4$\,rad\,m$^{-2}$), we aligned
the profiles by adjusting both phase and PA in all three frequencies,
obtaining a more precise value $\mbox{RM} = +77 \pm 1$\,rad\,m$^{-2}$.
Longitude 0\arcdeg\ on the plot is arbitrary.
}
\end{center}
\end{figure}

With the variability of the integrated profiles as a caveat, we
nevertheless attempt to compare the profiles at 1.4, 3.2 and 8.4\,GHz.
In order to isolate long-term variations, we consider for this purpose
only data taken in a one-week period in 2006 September.  The double
peaked profile gets narrower as the frequency increases and the ratio of
the leading to trailing component becomes larger (Fig.~\ref{fig:pa}).
Also, the slow PA sweep (and absolute value of the PA) is identical
at all frequencies, as expected in the ``rotating vector model''
of \citet{rc69a}.  This is consistent with the radius-to-frequency
mapping paradigm in which lower frequencies are emitted farther from the
star than higher frequencies \citep[e.g.,][]{cor78}.  Without detailed
geometrical information, however, it is difficult to quantify this effect
\citep[for a brief discussion of radius-to-frequency mapping concerning
\xte, see][]{crh+06,drr07}.

\begin{figure}
\begin{center}
\includegraphics[angle=270,scale=0.54]{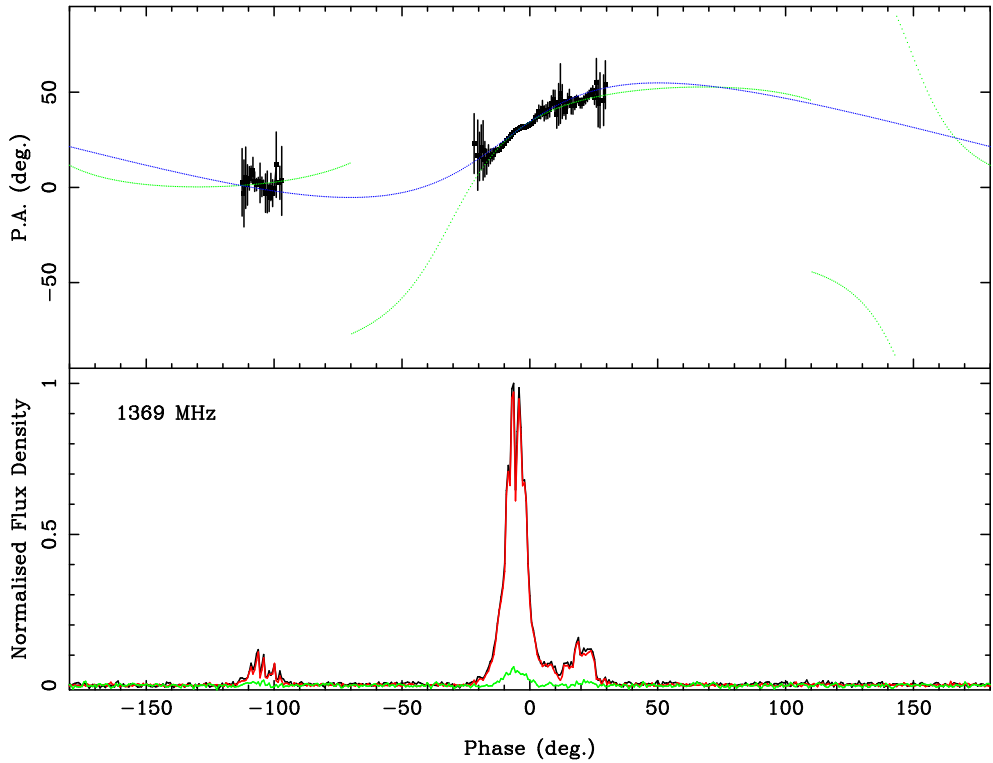}
\caption{\label{fig:pafit}
\textit{Bottom panel}: Polarimetric profile of \xte\ at 1.4\,GHz based
on 17\,min of data from MJD~53913.  The full 360\arcdeg\ of pulse phase
are shown.  The \textit{black} trace shows the total intensity emission,
while \textit{red} displays linear polarization and \textit{green}
represents the level of circular polarization.  \textit{Top panel}:
Position angle data (\textit{black} points) and the best-fit rotating
vector model curves for the almost aligned (\textit{blue} line) and
almost orthogonal (\textit{green} line) cases discussed in the text.
In the latter instance the emission from the main pulse component and
the precursor component must arise from different orthogonal modes,
thus necessitating the 90\arcdeg\ phase jumps in the fitted curve.
}
\end{center}
\end{figure}

In the very early days of pulsar astronomy, it was realized that
the observed PA swing could be used to derive the geometry of the
star under the assumption that the PA was related to the projection
of the dipolar field lines on the plane of the sky \citep{rc69a}.
Unfortunately, it is difficult to determine the geometry in the majority
of pulsars mainly because of the small longitude range over which they
emit \citep[e.g.,][]{ew01}.  This is true of \xte\ also.  In addition,
it is a priori unclear whether a dipolar field structure holds true in
this pulsar, although we proceed on the assumption that it might and
see where that leads us.  In our post-2006 July data, neither $\alpha$
(the angle between the magnetic and rotation axes) nor $\beta$ (the
angle of closest approach of the line of sight to the magnetic axis)
can be constrained.  The earlier data, with the appearance of pulse
components far from the main component (Fig.~\ref{fig:sp}), are more
promising in this regard.  Here, however, the main uncertainty is whether
there is 90\arcdeg\ of PA rotation between the widely spaced components.
Formal fits to the data both with and without an extra 90\arcdeg\ of
phase are reasonable (see Fig.~\ref{fig:pafit}).  In the former case,
the fits yield values of $\alpha$ near 70\arcdeg\ and high values of
$\beta$ near 20\arcdeg--25\arcdeg.  Without the added orthogonal jump,
the fits yield $\alpha \sim 4\arcdeg$ and $\beta \sim 4\arcdeg$ implying
that the magnetic and rotation axes are almost aligned.

The polarization characteristics of \xte\ are very similar to those
seen in young pulsar profiles \citep{jw06}. They too are highly linearly
polarized, often have double profiles, and show a slow swing of PA across
a wide profile.  \cite{jw06} showed that a single cone of emission
originating from relatively high in the pulsar magnetosphere could
explain the observed characteristics of young pulsars and it is tempting
to make the same case here. However, there is a significant difference
in the polar cap radius, $\propto P^{-1/2}$ (and light cylinder radius
$c P/2\pi$), between a young pulsar with a period of 0.1\,s and \xte\
with its 5.5\,s period.  This makes it difficult to see how such a wide
($\approx 0.15\,P$) observed pulse profile can be produced unless (a)
the emission height is very large or (b) the magnetic and rotation axes
are almost aligned.  We will discuss these two possibilities in turn.

In the first case, knowledge of the geometry and the observed pulse
width can be used to compute an emission height.  For values of $\alpha$
near 70\arcdeg\ and $\beta \approx 25\arcdeg$, one can use equation~(2)
of \citet{ggr84} to derive the cone opening angle $\rho$.  In turn the
emission height can be computed as $\sim 2cP \rho^2 / 9\pi$, or $\sim
20000$\,km.  This is about 10\% of the light cylinder radius --- similar
to the value in other young pulsars \citep{jw06}.  If this scenario
were typical of magnetars in general then the beaming fraction would be
high and most bright radio active magnetars would likely be detectable
in pulsations.  In the second case, for small values of $\alpha$, the
line of sight could remain wholly within the emission beam leading to the
observed wide profile.  In this case, if $\alpha$ were to vary slightly
with time (for reasons unknown), there could be a large effect on both the
observed beam shape and torque, both of which have been observed to vary
significantly \citep{ccr+07}.  Perhaps the emission in this particular
magnetar could be in part a direct function of the quasi-alignment
between the rotation axis and magnetic axis, or perhaps the alignment
might occur as a natural process in magnetars.  In either case, the small
polar cap size would make the beaming fraction of such magnetars rather
low.  The main difficulty with this interpretation is that the radio and
X-ray beams appear to be nearly aligned \citep{ccr+07} and the observed
modulation of thermal X-rays is very large \citep[$\sim 50\%$;][]{gh06},
which would be hard to obtain from a nearly aligned rotator.

In summary, the polarized emission from \xte\ shares many characteristics
of those in young pulsars generally. The emission is highly linearly
polarized with little evolution with frequency, the pulse profile
is wide and double, and there is only a shallow swing of PA through
the main pulse. This leads to the possibility that the ``standard''
pulsar ideas of emission along open magnetic field lines also hold here.
In this case, either the magnetic and rotation axes are almost aligned,
or the emission originates high above the surface of the star, which
is our preferred interpretation.  Obvious remaining differences between
\xte\ and other pulsars are its pulse profile variability (which does not
appear to be accompanied by corresponding gross changes in the magnetic
field geometry), fluctuating flux density, and flat spectrum.

\acknowledgements

We thank John Sarkissian for help with observations, and Aidan Hotan
and Aris Karastergiou for discussions.  The Parkes Observatory is
part of the Australia Telescope, which is funded by the Commonwealth
of Australia for operation as a National Facility managed by CSIRO.
FC acknowledges the NSF for support through grant AST-05-07376.

\begin{thebibliography}{20}
\expandafter\ifx\csname natexlab\endcsname\relax\def\natexlab#1{#1}\fi

\bibitem[{Beloborodov \& Thompson(2007)}]{bt07}
Beloborodov, A.~M., \& Thompson, C. 2007, ApJ, in press (astro-ph/0602417)

\bibitem[{{Camilo} {et~al.}(2007){Camilo}, Cognard, {Ransom}, {Halpern},
  {Reynolds}, {Zimmerman}, Gotthelf, Helfand, Demorest, Theureau, \&
  Backer}]{ccr+07}
{Camilo}, F., et~al. 2007, ApJ, submitted (astro-ph/0610685)

\bibitem[{{Camilo} {et~al.}(2006){Camilo}, {Ransom}, {Halpern}, {Reynolds},
  {Helfand}, {Zimmerman}, \& {Sarkissian}}]{crh+06}
{Camilo}, F., {Ransom}, S.~M., {Halpern}, J.~P., {Reynolds}, J., {Helfand},
  D.~J., {Zimmerman}, N., \& {Sarkissian}, J. 2006, Nature, 442, 892

\bibitem[{Cordes(1978)}]{cor78}
Cordes, J.~M. 1978, ApJ, 222, 1006

\bibitem[{Duncan \& Thompson(1992)}]{dt92a}
Duncan, R.~C., \& Thompson, C. 1992, ApJ, 392, L9

\bibitem[{{Dyks} {et~al.}(2007){Dyks}, {Rudak}, \& {Rankin}}]{drr07}
{Dyks}, J., {Rudak}, B., \& {Rankin}, J.~M. 2007, A\&A, in press
  (astro-ph/0610883)

\bibitem[{{Everett} \& {Weisberg}(2001)}]{ew01}
{Everett}, J.~E., \& {Weisberg}, J.~M. 2001, ApJ, 553, 341

\bibitem[{Gil {et~al.}(1984)Gil, Gronkowski, \& Rudnicki}]{ggr84}
Gil, J.~A., Gronkowski, P., \& Rudnicki, W. 1984, A\&A, 132, 312

\bibitem[{Gotthelf \& Halpern(2005)}]{gh05}
Gotthelf, E.~V., \& Halpern, J.~P. 2005, ApJ, 632, 1075

\bibitem[{{Gotthelf} \& {Halpern}(2007)}]{gh06}
{Gotthelf}, E.~V., \& {Halpern}, J.~P. 2007, in Isolated Neutron Stars: From
  the Interior to the Surface, ed. S.~Zane, R.~Turolla, \& D.~Page, in press
  (astro-ph/0608473)

\bibitem[{Halpern {et~al.}(2005)Halpern, Gotthelf, Becker, Helfand, \&
  White}]{hgb+05}
Halpern, J.~P., Gotthelf, E.~V., Becker, R.~H., Helfand, D.~J., \& White, R.~L.
  2005, ApJ, 632, L29

\bibitem[{{Han} {et~al.}(2006){Han}, {Manchester}, {Lyne}, {Qiao}, \& {van
  Straten}}]{hml+06}
{Han}, J.~L., {Manchester}, R.~N., {Lyne}, A.~G., {Qiao}, G.~J., \& {van
  Straten}, W. 2006, \apj, 642, 868

\bibitem[{{Hotan} {et~al.}(2004){Hotan}, {van Straten}, \&
  {Manchester}}]{hvm04}
{Hotan}, A.~W., {van Straten}, W., \& {Manchester}, R.~N. 2004, Proc. Astr.
  Soc. Aust., 21, 302

\bibitem[{{Ibrahim} {et~al.}(2004){Ibrahim}, {Markwardt}, {Swank}, {Ransom},
  {Roberts}, {Kaspi}, {Woods}, {Safi-Harb}, {Balman}, {Parke}, {Kouveliotou},
  {Hurley}, \& {Cline}}]{ims+04}
{Ibrahim}, A.~I., et~al. 2004, ApJ, 609, L21

\bibitem[{Johnston(2002)}]{joh02}
Johnston, S. 2002, Proc. Astr. Soc. Aust., 19, 277

\bibitem[{{Johnston} {et~al.}(2005){Johnston}, {Hobbs}, {Vigeland}, {Kramer},
  {Weisberg}, \& {Lyne}}]{jhv+05}
{Johnston}, S., {Hobbs}, G., {Vigeland}, S., {Kramer}, M., {Weisberg}, J.~M.,
  \& {Lyne}, A.~G. 2005, \mnras, 364, 1397

\bibitem[{{Johnston} \& {Weisberg}(2006)}]{jw06}
{Johnston}, S., \& {Weisberg}, J.~M. 2006, \mnras, 368, 1856

\bibitem[{{Karastergiou} \& {Johnston}(2006)}]{kj06}
{Karastergiou}, A., \& {Johnston}, S. 2006, \mnras, 365, 353

\bibitem[{Radhakrishnan \& Cooke(1969)}]{rc69a}
Radhakrishnan, V., \& Cooke, D.~J. 1969, Astrophys. Lett., 3, 225

\bibitem[{Woods \& Thompson(2006)}]{wt06}
Woods, P.~M., \& Thompson, C. 2006, in Compact Stellar X-ray Sources, ed.
  W.~H.~G. Lewin \& M.~van~der Klis (Cambridge: CUP), 547--586

\end{thebibliography}
\end{document}